\documentclass[conference,twoside,10pt]{IEEEtran}
\pdfoutput=1
\usepackage{graphicx,color}
\usepackage[dvipsnames]{xcolor}
\usepackage{amsmath,amsbsy,amsfonts,amssymb}
\usepackage{mathrsfs}
\usepackage{mathtools}
\mathtoolsset{showonlyrefs}
\usepackage{csquotes}
\ifCLASSOPTIONcompsoc
\usepackage[caption=false,font=normalsize,labelfont=sf,textfont=sf]{subfig}
\else
\usepackage[caption=false,font=footnotesize]{subfig}
\fi
\usepackage[noadjust]{cite}
\IEEEoverridecommandlockouts

\usepackage{acro}
\DeclareAcronym{AWGN}{short = AWGN ,long = additive white gaussian noise}
\DeclareAcronym{AoI}{short = AoI ,long = age of information}
\DeclareAcronym{PAoI}{short = PAoI ,long = peak age of information}
\DeclareAcronym{AT-IRSA}{short = AT-IRSA ,long = age-threshold IRSA}
\DeclareAcronym{CDF}{short = CDF ,long = cumulative distribution function}
\DeclareAcronym{CRA}{short = CRA ,long = contention resolution ALOHA}
\DeclareAcronym{CRDSA}{short = CRDSA ,long = contention resolution diversity slotted ALOHA}
\DeclareAcronym{CP}{short = CP ,long = contention period}
\DeclareAcronym{CSA}{short = CSA ,long = coded slotted ALOHA}
\DeclareAcronym{C-RAN}{short = C-RAN ,long = cloud radio access network}
\DeclareAcronym{DAMA}{short = DAMA ,long = demand assigned multiple access}
\DeclareAcronym{DSA}{short = DSA ,long = diversity slotted ALOHA}
\DeclareAcronym{eMBB}{short = eMBB ,long = enhanced mobile broadband}
\DeclareAcronym{FEC}{short = FEC ,long = forward error correction}
\DeclareAcronym{GEO}{short = GEO ,long = geostationary orbit}
\DeclareAcronym{HAP}{short = HAP ,long = high-altitude platform,foreign-plural={}}
\DeclareAcronym{IC}{short = IC ,long = interference cancellation}
\DeclareAcronym{IoT}{short = IoT ,long = Internet of things}
\DeclareAcronym{IRSA}{short = IRSA ,long = irregular repetition slotted ALOHA}
\DeclareAcronym{LEO}{short = LEO ,long = low Earth orbit}
\DeclareAcronym{M2M}{short = M2M ,long = machine-to-machine}
\DeclareAcronym{MAC}{short = MAC ,long = medium access}
\DeclareAcronym{MPR}{short = MPR ,long = multi-packet reception}
\DeclareAcronym{MTC}{short = MTC ,long = machine-type communications}
\DeclareAcronym{mMTC}{short = mMTC ,long = massive machine-type communications}
\DeclareAcronym{NTN}{short = NTN ,long = non-terrestrial network,foreign-plural = {}}
\DeclareAcronym{PDF}{short = PDF ,long = probability density function}
\DeclareAcronym{PER}{short = PER ,long = packet error rate}
\DeclareAcronym{PLR}{short = PLR ,long = packet loss rate}
\DeclareAcronym{PMF}{short = PMF ,long = probability mass function}
\DeclareAcronym{RA}{short = RA ,long = random access}
\DeclareAcronym{RRH}{short = RRH ,long = remote radio head,foreign-plural = {}}
\DeclareAcronym{rv}{short = r.v. ,long = random variable,foreign-plural = {}}
\DeclareAcronym{SA}{short = SA , long = slotted ALOHA}
\DeclareAcronym{SIC}{short = SIC ,long = successive interference cancellation}
\DeclareAcronym{SIR}{short = SIR ,long = signal to interference ratio}
\DeclareAcronym{SNIR}{short = SNIR ,long = signal-to-noise and interference ratio}
\DeclareAcronym{SINR}{short = SINR ,long = signal-to-interference and noise ratio}
\DeclareAcronym{SNR}{short = SNR ,long = signal-to-noise ratio}
\DeclareAcronym{TDM}{short = TDM ,long = time division multiplexing}

\newtheorem{remark}{Remark}

\DeclareMathOperator*{\argmax}{argmax}   

\newcommand{\expOp}{\ensuremath{\mathbb E}}

\newcommand{\tru}{\ensuremath{\mathsf S}}

\newcommand{\load}{\ensuremath{\mathsf G}}
\newcommand{\maxLoad}{\ensuremath{\mathsf G^*}}

\newcommand{\psucc}{\ensuremath{\mathsf p_{s}}}

\newcommand{\nodes}{\ensuremath{\mathsf U}}
\newcommand{\nodesRV}{\ensuremath{U_\ell}}
\newcommand{\loadRV}{\ensuremath{G_\ell}}
\newcommand{\slots}{\ensuremath{\mathsf m}}
\newcommand{\age}{\ensuremath{\delta}}
\newcommand{\Age}{\ensuremath{\Delta}}
\newcommand{\thr}{\ensuremath{\Theta}}

\begin{document}

\title{\huge On the Performance of Irregular Repetition Slotted ALOHA with an Age of Information Threshold}
\author{
\IEEEauthorblockN{Hooman Asgari}
\IEEEauthorblockA{
\textit{Technical University of Munich}\\
Munich, Germany \\
hooman.asgari@tum.de}
\and
\IEEEauthorblockN{Andrea Munari}
\IEEEauthorblockA{
\textit{German Aerospace Center (DLR)}\\
Wessling, Germany \\
andrea.munari@dlr.de}
\and
\IEEEauthorblockN{Gianluigi Liva}
\IEEEauthorblockA{
\textit{German Aerospace Center (DLR)}\\
Wessling, Germany \\
gianluigi.liva@dlr.de}
\thanks{A. Munari and G. Liva acknowledge the financial support by the Federal Ministry of Education and Research of Germany in the programme of "Souver\"an. Digital. Vernetzt." Joint project 6G-RIC, project identification number: 16KISK022.
}
}

\maketitle
\thispagestyle{empty} \setcounter{page}{0}

\maketitle

\pagestyle{empty}

\begin{abstract}
  The present paper focuses on an IoT setting in which a large number of devices generate time-stamped updates addressed to a common gateway. Medium access is regulated following a grant-free approach, and the system aims at maintaining an up-to-date knowledge at the receiver, measured through the average network age of information (AoI). In this context, we propose a variation of the irregular repetition slotted ALOHA (IRSA) protocol. The scheme, referred to as age-threshold IRSA (AT-IRSA), leverages feedback provided by the receiver to admit to the channel only devices whose AoI exceeds a dynamically adapted target value. By means of detailed networks simulations, as well as of a simple yet tight analytical approximation, we demonstrate that the approach can more than halve the average network AoI compared to plain IRSA, and offers notable improvements over feedback-based state-of-the-art slotted ALOHA solutions recently proposed in the literature.
\end{abstract}

\section{Introduction}
\label{sec:intro}

Massive connectivity for the \ac{IoT} is attracting a steadily increasing attention, and is envisaged to become a key component of upcoming 6G wireless systems. The possibility to gather data from a large population of low-power, low-battery devices that generate traffic in a sporadic manner paves the way to a number of applications, ranging among others from environmental and industrial monitoring to asset tracking. In many of these settings, the ultimate goal is to maintain an up-to-date perception of some monitored processes at one or more collection points, e.g. to trigger appropriate actuation steps.

Motivated by this remark, recent research efforts have focused on the definition of strategies, often labeled as \emph{semantic communications}, that aim to facilitate delivery of the right piece of information to the right point in time for \ac{IoT} applications \cite{Uysal22_Semantic}. To capture this ability, a number of new performance metrics were introduced, ranging from \ac{AoI} \cite{Yates19_TIT} to more advanced indicators such as the age of incorrect information \cite{Ephremides20:TNET}, the value of information \cite{Soleymani20_valueInfo,Kellerer19}, or the query AoI \cite{Chiariotti22:TCOM,Uysal22:ISIT}.

Among these, \ac{AoI} had a pioneering role. Originally proposed for vehicular networks \cite{Kaul11_Secon,Kaul11_Globecom}, the metric quantifies the time elapsed since the generation of the freshest available update on a tracked process. In spite of its simplicity, \ac{AoI} has been shown to be an effective proxy to gauge the fundamental trade-offs in a number of \ac{IoT} and cyber-physical systems \cite{Kellerer19,Uysal20_TIT}, and as such is of particular relevance. A good level of maturity has been reached for the \ac{AoI} behavior in point-to-point communication links, see e.g. \cite{Uysal20_TIT,Durisi19_JSAC} as well as \cite{Yates19_TIT} and references therein. On the other hand, only recently studies have started to focus on the implications of information freshness in multi-access systems, spanning different layers of the protocol stack \cite{Shreedhar22:INFOCOM,Uysal21:ACP,Modiano19_TNET,Ephremides19_Infocom}.


From this standpoint, channel access strategies are of particular interest in the context of \ac{IoT} connectivity. In fact, the intermittent activity of a massive number of devices renders traditional grant-based link-layer solutions ineffective, and  random access strategies based on variations of the ALOHA paradigm \cite{Abramson77:PacketBroadcasting} are commonly employed in practical systems \cite{LoRa}. First fundamental results on the performance of such schemes in terms of \ac{AoI} have been obtained \cite{Yates17:AoI_SA,Modiano18_AoI,Yates20_ISIT,Munari22:Globecom,Munari21:Balkancom}, highlighting some specific design and optimization criteria to target information freshness. In parallel, research has focused on the study on a family of grant-free protocols designed in the early 2000s, capable to go beyond the intrinsic throughput and reliability limitations of the collision-beset ALOHA approach \cite{Paolini14_CommMag}. These solutions, often dubbed \emph{modern random access}, allow nodes to proactively transmit multiple copies of a packet, relying on successive interference cancellation at the receiver side to resolve collisions, and have been shown to approach performance competitive with that of coordinated access  \cite{Paolini15:TIT,Stefanovic12:COML,Sandgren17_TCOM,Clazzer18:ECRA}. Remarkable \ac{AoI} improvements have also been recently demonstrated for such schemes \cite{Munari21_TCOM_AoI,Munari21:Asilomar,Ngo21:Asilomar}, prompting them as promising candidates for \ac{IoT} settings.

These lines of research typically focus on random access settings that do not rely on feedback, assuming devices to transmit without knowledge of the current AoI at the receiver. This hypothesis was relaxed by some notable works in the past few years, which showed how the availability of a return channel can dramatically improve performance in slotted ALOHA \cite{Uysal21_AlohaThresh,Uysal22:MiSTA,Bidokhti22:TIT}. In particular, the possibility to prioritize channel access for nodes perceived by the sink as stale by means of thresholding policies has proven capable to approximately halve the average \ac{AoI} with negligible losses in terms of throughput.

The impact of feedback for the AoI of advanced grant-free schemes remains to date instead largely unexplored. In this paper we start to bridge this gap focusing on a modern random access protocol, namely \ac{IRSA}  \cite{Liva11:TCOM}. We discuss the implications and costs of implementing a feedback for such scheme, and propose a variation of the algorithm, referred to as \ac{AT-IRSA}, which lets the receiver dynamically select a minimum level of \ac{AoI} required for nodes to contend. By means of detailed network simulations we prove the remarkable benefits of such approach compared to the basic version of \ac{IRSA}, especially for large terminal populations. Furthermore, we present a simple yet tight analytical approximation for the performance of the scheme, and show that substantial improvements over some state-of-the-art feedback-based slotted ALOHA policies \cite{Bidokhti22:TIT,Uysal21_AlohaThresh} are attained.

\section{System Model and Preliminaries}
\label{sec:sysModel}

Throughout our discussion we focus on a system composed by a large number of terminals which share a common wireless channel to communicate to a common receiver (also referred to as sink, or monitor). Each of the \nodes\ devices becomes active sporadically, attempting transmission of a time-stamped status update, e.g., containing the reading of a monitored physical process. In this setting, we aim to maintain a fresh and up-to-date perception of the monitored processes at the receiver.

As customary in massive \ac{IoT} applications, a random access approach is implemented at the link layer, following a variation of the \acl{IRSA} protocol (IRSA) described in details in Sec.~\ref{sec:IRSA} and \ref{sec:AT-IRSA}. In the remainder of our discussion, we assume time to be divided in slots of equal duration, and all devices to be synchronized to such pattern. The transmission parameters are set such that a slot fits a single packet. Furthermore, following a well-established modeling approach, we regard collisions as destructive. Accordingly, the receiver cannot extract any information from a slot containing the superposition of two or more packets. Conversely, a data unit without interference (singleton slot) is always correctly decoded. In addition, we assume the sink to be able to differentiate among idle, singleton and collided slots.

\begin{figure*}[ht!]
  \centering
  \subfloat[]{\includegraphics[width=.24\textwidth]{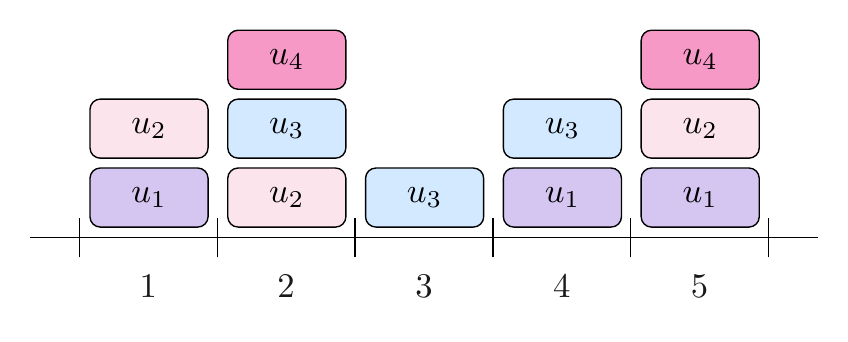}\label{fig:irsa_A}}
  \subfloat[]{\includegraphics[width=.24\textwidth]{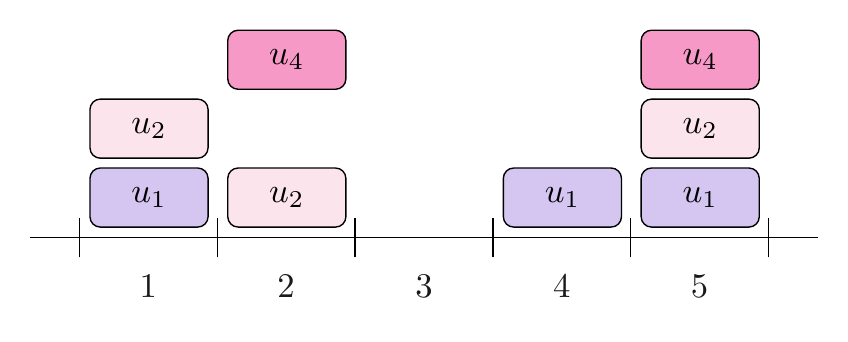}\label{fig:irsa_B}}
  \subfloat[]{\includegraphics[width=.24\textwidth]{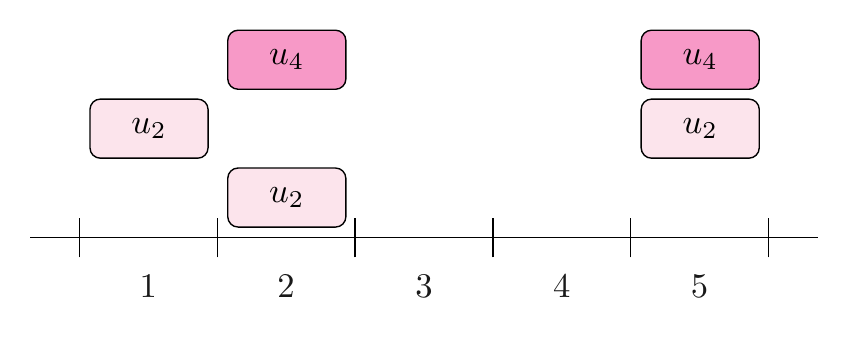}\label{fig:irsa_C}}
  \subfloat[]{\includegraphics[width=.24\textwidth]{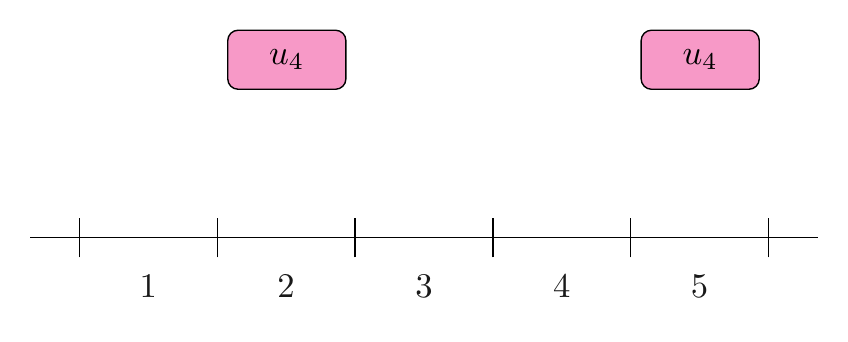}\label{fig:irsa_D}}
  \caption{Example timeline for operation of IRSA at the receiver side. In the considered case, $4$ users access a frame of $\slots=5$ slots. A complete discussion is provided in Sec.~\ref{sec:IRSA}}.
  \label{fig:irsa_timeline}
\end{figure*}

\subsection{Irregular Repetition Slotted ALOHA}
\label{sec:IRSA}
Originally introduced in \cite{Liva11:TCOM}, \ac{IRSA} is a grant-free scheme designed to go beyond the intrinsic reliability and throughput limitations of slotted ALOHA. The protocol operates over frames of \slots\ slots each,\footnote{Variations of the protocol operating over group of resources allocated in a time-frequency thread are also possible, see, e.g. \cite{dvbrcs2}.} and the delivery of a packet can only be initiated at the start over a new frame. Specifically, when a terminal has data to send, it will transmit $\ell$
 copies of the packet, uniformly distributed at random over the \slots\ available slots in the upcoming frame. Each replica contains a pointer to the positions in which its twins are sent.\footnote{This can be implemented by signaling the slots over which transmissions are performed in the packet header. Alternative solutions to reduce overhead are also possible, e.g. using the payload as seed for a random number generator,
used both at the sender and receiver side to place and locate replicas.} The number of copies is drawn from a pre-defined distribution, shared by all devices in the network. Following a well-established notation, we specify such distribution in polynomial form as
 \begin{equation}
   \Lambda(x) = \sum_{\ell=1}^{\mathsf L} \Lambda_\ell \, x^\ell
 \end{equation}
where $\Lambda_\ell$ is the probability to send $\ell$ packet copies, up to a maximum degree $\mathsf L$.

\begin{figure}
  \centering
  \includegraphics[width=.8\columnwidth]{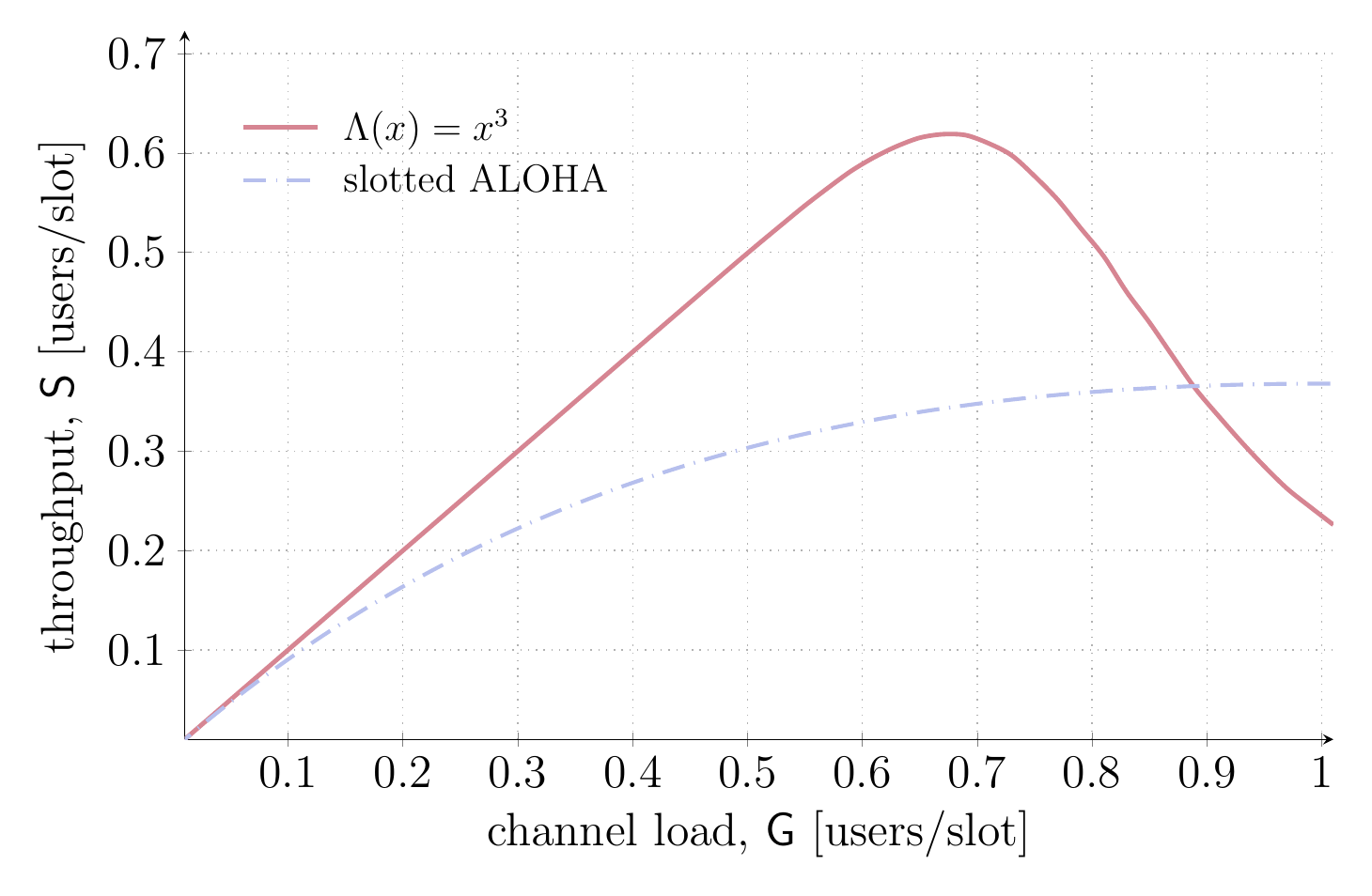}
  \caption{Throughput vs channel load, simulation results. The solid line reports the behavior of IRSA with distribution $\Lambda(x)=x^3$, operated over frames of duration $\slots=100$ slots. The dashed line shows the well-known performance of slotted ALOHA. All results obtained for $\nodes=4000$ terminals.}
  \label{fig:truVsLoad}
\end{figure}

At the sink, decoding relies on \ac{SIC} procedures. After buffering a whole frame, the receiver starts by identifying singleton slots, which allow retrieval of non-collided packets. The incoming signal contribution of each decoded data unit is then subtracted from all the slots in which its copies were transmitted. This interference cancellation procedure can thus potentially lead to the identification of additional singleton slots, and is iterated until either all packets have been decoded or only slots with collisions remain in the frame. An example of the described receiver operation is reported in Fig.~\ref{fig:irsa_timeline}. In this case, $4$ users access a frame of duration $5$ slots. Users $1$, $2$ and $3$ transmit three copies of their packet, whereas user $4$ only sends two replicas, leading to the initial configuration of Fig.~\ref{fig:irsa_A}. The receiver starts by decoding user $3$ from slot $3$ (singleton), and removes the contribution of such packet from slots $2$ and $4$ (Fig.~\ref{fig:irsa_B}). At this point, slot $4$ only contains the packet of user $1$, which can be retrieved. Once more, the corresponding signal is canceled from slots $1$ and $5$, obtaining the configuration of Fig.~\ref{fig:irsa_C}.
Here, user $2$ is decoded from the first slot, eventually resolving all collisions involving user $4$ as well (Fig.\ref{fig:irsa_D}).

To characterize the performance of \ac{IRSA} in the remainder of our discussion we resort to two key figures: channel load and throughput. The former, denoted by \load, captures the level of contention over a frame. More precisely, let us introduce the \ac{rv}
\begin{equation}
\loadRV := \frac{\nodesRV}{\slots}
\end{equation}
describing the instantaneous channel load over the $\ell$-th frame, where the r.v. $\nodesRV$ indicates the number of terminals attempting a transmission. The average channel load is accordingly
\begin{equation}
  \load = \expOp\left[\loadRV\right]\,.
\end{equation}
In turn, the throughput \tru\ is defined as the average number of terminals decoded per slot 
The throughput behavior of \ac{IRSA} has been thoroughly studied in the literature, see e.g., \cite{Liva11:TCOM,Paolini15:TIT}, and an example of the achievable performance is reported in Fig.~\ref{fig:truVsLoad} for a frame size $\slots=100$ and an overall population of $\nodes=4000$ users.

\subsection{Age of Information: Preliminaries}

To gauge the ability of a random access policy to maintain a fresh perception of monitored processes at the sink, we consider the \acl{AoI} metric. Focusing without loss of generality on an arbitrary terminal $u$ in the system, let us denote as $\age_u(t)$ its  instantaneous \ac{AoI}, defined as
\begin{equation}
  \age_u(t) := t - \sigma_u(t)
\end{equation}
where $\sigma(t)$ is the time-stamp of the last successfully received update from the node as of time $t$. Leaning on this, we introduce the average \ac{AoI} for the node
\begin{equation}
  \Age_u := \lim_{t\rightarrow \infty} \int_{0}^t \age_u(\tau) \,d\tau
\end{equation}
assuming that the limit exists, and study in the remainder the \emph{average network} \ac{AoI}
\begin{equation}
  \Age = \frac{1}{\nodes} \sum_{u=1}^\nodes \Age_u \,.
\end{equation}

\begin{figure}
  \centering
  \includegraphics[width=.75\columnwidth]{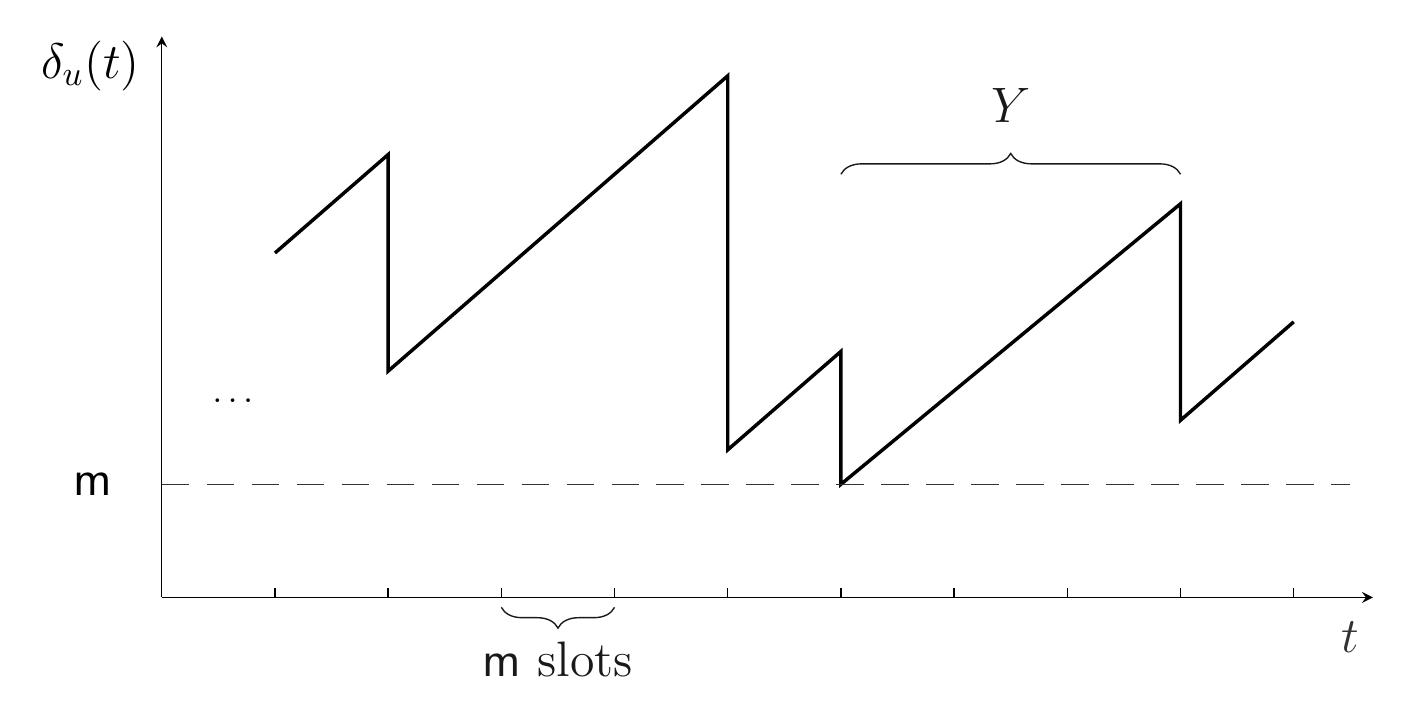}
  \caption{Example timeline for the evolution of the instantaneous AoI of a node. The inter-update time $Y$ is also reported.}
  \label{fig:aoi_timeline}
\end{figure}

When the system is operated following the \ac{IRSA} access policy we assume that all copies of a packet sent by a terminal contain a time-stamp marked with the start of the frame. This corresponds to the commonly employed \emph{generate-at-will} model, capturing the possibility for a device to perform a reading of the process being monitored right before sending the status report. Moreover, in case of successful decoding, the instantaneous AoI for a terminal is reset to a value of $\slots$ slots, accounting for the fact that the receiver buffers the whole frame before triggering \ac{SIC} (and neglecting the processing time). Following these remarks, the \ac{AoI} evolution of a device follows the sawtooth profile exemplified in Fig.~\ref{fig:aoi_timeline}. Under these assumptions, an exact characterization of the average network AoI for IRSA can be obtained by simply adapting the results in \cite{Munari21_TCOM_AoI}, to obtain
\begin{equation}
  \Age_{\textsf{irsa}} = \frac{\slots}{2} + \frac{\nodes}{\tru}
  \label{eq:age_irsa}
\end{equation}
Note that, for a given frame size \slots\, the \ac{AoI} is minimized under maximum throughput operations.

\section{Age-Threshold IRSA}
\label{sec:AT-IRSA}

The original design and optimization of \ac{IRSA} \cite{Liva11:TCOM,Paolini15:TIT}, as well as existing studies of its  \ac{AoI} performance \cite{Munari21_TCOM_AoI,Ngo21:Asilomar}, were carried out considering an i.i.d. activation profile of terminals across frames. This assumption can be practical in many \ac{IoT} settings where devices monitor uncorrelated processes. On the other hand, when the aim of the system is to maintain fresh information at the sink, prioritizing transmissions of terminals with higher instantaneous \ac{AoI} may be more beneficial. Taking the lead from this observation, we propose a modification of the access protocol, named age-threshold IRSA (AT-IRSA), capable to lower the average network \ac{AoI} while preserving throughput performance. Before describing the algorithm in Sec.~\ref{sec:protocol}, some preliminary remarks are in order.

\begin{remark}{\emph{Feedback availability.}}
In the remainder, we assume the sink to broadcast at the end of each frame an error-free feedback to all terminals that attempted transmission, specifying which packets have been decoded successfully. We observe that this operating condition is reasonable, as a forward link is typically available in practical implementations of IRSA (e.g. \cite{dvbrcs2}) for the purpose of maintaining synchronization among terminals. From this viewpoint, the feedback could be simply piggybacked onto a beacon signaling the start of a new frame. It shall also be noted that the overhead cost for implementing the feedback is expected to be small. For example, signaling the binary outcome (completely resolved or residual collision) of each slot can allow all active nodes to understand whether their updates were retrieved. Accordingly, \slots\ bits would suffice to convey the return link message. For a typical duration of an \ac{IRSA} frame, the feedback would then entail a cost of few hundred bits, comparable to one or few uplink packets at most. We also remark that the assumption of  providing feedback at the end of a frame can be less demanding and more appealing in practical systems compared to the implementation of a receiver response on a slot basis required for operation in slotted ALOHA variations.
\end{remark}

\begin{remark}{\emph{Age-threshold approach.}}
Through the availability of feedback, each terminal in the network can track its current \ac{AoI} as seen at the receiver. Leaning on this, \Age\ can be reduced by favoring delivery attempts for processes affected by a more stale perception at the monitor, rather than granting the same probability of access to all terminals regardless of their status. A simple policy to implement this approach consists in defining a threshold \thr, and allowing only terminals with instantaneous \ac{AoI} above it to access the channel over a frame with a certain probability. From this standpoint, several criteria can be identified to specify the value of \thr. Among them, we focus on the possibility to maintain the instantaneous channel load around the value achieving peak throughput. The choice is meant to maximize the number of updates per slot that can be delivered in the system, and is consistent with the inverse proportionality of the average \ac{AoI} to throughput exhibited by \ac{IRSA}, see \eqref{eq:age_irsa}. Incidentally, we note that this strategy leads to a dynamic adaptation of the threshold, increasing it when a larger fraction of the available observations are stale, and lowering it when devices with smaller instantaneous \ac{AoI} shall be granted access to the channel.
\end{remark}

\subsection{AT-IRSA operations}
\label{sec:protocol}

In order to introduce the proposed algorithm, let us denote by \maxLoad\ the average channel load for which the aggregate throughput \tru\ is maximized for the considered frame length \slots. Moreover, for any $\theta \in \mathbb R^+$, let $\mathcal U(\theta)$ be the set of terminals whose current instantaneous \ac{AoI} is larger than $\theta$, and $\mathsf n(\theta)$ the cardinality of such set.

Leaning on this notation, the transmission and reception operations of the protocol follow the regular \ac{IRSA} behavior, with the following modifications. At the end of each frame, after completion of the \ac{SIC} procedures, the receiver shall:
\begin{itemize}
    \item determine the age threshold \thr\ to be employed over the subsequent frame, computed as
      \begin{align}
        \thr =& \argmax_{\theta} \mathsf n(\theta)\\
        \text{s.t.} & \quad \mathsf n(\theta) \geq \slots \maxLoad
      \end{align}
    \item compute a barring probability, given by
      \begin{equation}
        p = \frac{\slots\maxLoad}{\mathsf n(\thr)}
      \end{equation}
    \item piggyback the pair $(\thr,p)$ to the feedback and broadcast the packet on the forward link.
\end{itemize}

At the terminal side, all devices listen for the incoming feedback, and update their current \ac{AoI} value accordingly, resetting it to \slots\ in case they delivered a packet successfully, or increasing it by \slots\ slot durations otherwise. After this, any terminal whose current \ac{AoI} is above the received value of \thr\ generates an update with probability $p$, transmitting it over the successive frame following the regular \ac{IRSA} procedures. All other nodes remain silent for the upcoming \slots\ slots.

\section{Results and Discussion}
\label{sec:results}

To evaluate the potential of the proposed approach, a set of detailed network simulations were performed, investigating the age performance of both IRSA and AT-IRSA. Throughout our study, we considered the regular distribution $\Lambda(x)=x^3$ for the number of transmitted replicas.

\begin{figure}
  \centering
  \includegraphics[width=0.85\columnwidth]{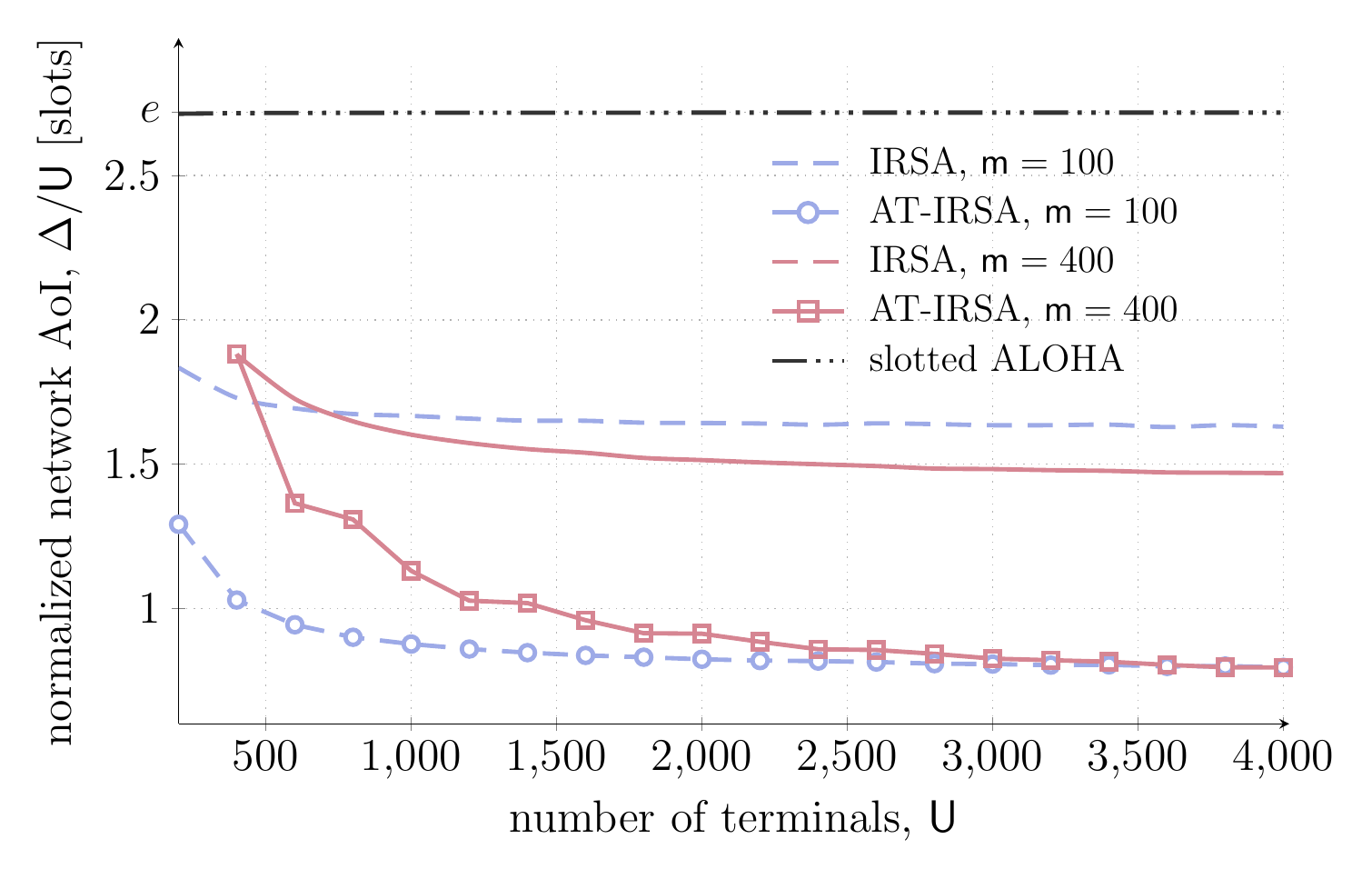}
  \caption{Normalized network AoI for \ac{AT-IRSA} vs number of terminals in the system, \nodes. For IRSA and AT-IRSA, $\Lambda(x)=x^3$ was employed.}
  \label{fig:basic_results}
\end{figure}

First interesting insights are offered in Fig.~\ref{fig:basic_results}, reporting the network AoI \Age\ normalized to the terminal population \nodes. Lines with no markers denote the performance of IRSA, whereas marked ones refer to its age-threshold variation. For both solutions, results are shown for a frame duration of $\slots=100$ and $\slots=400$ slots. In all cases, the system was operated at a target channel load that maximizes the throughput ($\load=0.66$ for $\slots=100$ and $\load=0.73$ for $\slots=400$). As discussed, for plain IRSA this corresponds to implementing an i.i.d. activation pattern for nodes across frames, so that the desired average number of transmissions is performed. For AT-IRSA, instead, the specificed load value is employed as parameter $\maxLoad$ in the algorithm.
For completeness, Fig.~\ref{fig:basic_results} also reports the performance of a basic slotted ALOHA solution. In this case, the average network AoI is given by $\Age_{\textsf{sa}} = 1/2 + \nodes/\tru$, see, e.g., \cite{Munari21_TCOM_AoI,Yates17:AoI_SA}. Results are once more obtained under optimal throughput (and minimum AoI) conditions, employing a channel access probability of $1/\nodes$, so that $\Delta_{\textsf{sa}}$ rapidly converges to $e \nodes$ as the network population grows \cite{Munari21_TCOM_AoI}.

As highlighted by the plot, the proposed solution attains remarkable improvements over IRSA, almost halving the network AoI for large values of $\nodes$ and confirming the potential of age-threshold based strategies. Two further remarks emerge from the figure. First, the performance of both schemes improves when increasing the number of terminals. This effect can be explained observing how for larger values of \nodes\ the fluctuations in the number of contending terminals around the targeted channel load value tends to reduce. Such a concentration has a beneficial impact on the average achieved throughput and reflects on the AoI statistics as well. Second, a better behavior in terms of information freshness can be achieved when operating the system over longer frames. The rationale behind the trend can be appreciated considering the \ac{AoI} formulation for IRSA reported in \eqref{eq:age_irsa}. The expression elegantly captures a fundamental trade-off between frame duration and average throughput. Indeed, while employing shorter frames reduces the first \ac{AoI} component  ($\slots/2$), it is well known that larger values of \slots\ allow to operate at higher throughput (lowering the $\nodes/\tru$ \ac{AoI} component) \cite{Paolini15:TIT,Munari21_TCOM_AoI}. From this standpoint, the second addend in \eqref{eq:age_irsa} becomes predominant as the network population increases, driving the trade-off in favor of longer frames.
Finally, Fig.~\ref{fig:basic_results} pinpoints the substantial \ac{AoI} reduction allowed by AT-IRSA in comparison to a baseline slotted ALOHA approach. For example, when serving $\nodes=4000$ terminals, the network AoI granted by the proposed solution is less than $1/3$ of the one undergone with the commonly employed ALOHA policy.

To gain further understanding on how the behavior of AT-IRSA scales with the network size, we complement our study with a tractable yet insightful analytical approximation. To this aim, let us focus on the example timeline reported in Fig.~\ref{fig:aoi_timeline}. Under the assumption of ergodicity for the instantaneous age process, it is well-known that the average network AoI can be characterized as (see e.g., \cite[Eq.~(3)]{Yates19_TIT})
\begin{equation}
  \Age = \frac{\expOp[M Y] + \expOp[Y^2/2]}{\expOp[Y]}
\end{equation}
where $Y$ is the \ac{rv} describing the inter-update times, i.e., the number of slots that elapse between reception of two successive updates from a terminal, and $M$ is the packet service time, i.e., the time between generation and reception of an update. In the setting under study, $M$ is a constant, equal to the frame duration, so that the expression simplifies to
\begin{align}
  \Age = \slots + \frac{\expOp[Y^2]}{2 \expOp[Y]}
  \label{eq:age_AT-IRSA}
\end{align}
and the AoI can be computed by evaluating the first and second order moment of the inter-update times. On the other hand, we remark that an exact characterization of $Y$ is in general not trivial for AT-IRSA. Indeed, the possibility for multiple users not to be decoded over the same frame, combined with the proposed age-threshold policy, triggers a correlation in the behavior of terminals over time. We disregard for the moment these aspects, and derive an approximation of the inter-update times as follows.

We start by observing that, after successfully delivering a packet, a reference terminal will have to wait for a certain number of frames before being allowed to transmit again (i.e. for its AoI to exceed the threshold). As a first approximation, we consider a round-robin like behavior: observing that on average $\slots \maxLoad$ nodes access a frame, the terminal will be allowed to transmit again in around $A := \nodes/(\slots \maxLoad)$ frames.\footnote{Note that, in the presented approximation, no dynamic adaptation of the threshold is considered.}
Once the age threshold is exceeded, the node accesss the channel in each of the successive frames with probability $p$ to attempt delivery. Accordingly, additional $B$ frames are needed to terminate the current inter-update period. Disregarding the impact of $p$, we assume for $B\geq 0$ a geometric distribution with parameter $\psucc$, where the \psucc\ denotes the probability of decoding a packet sent over a frame. Following this approach, we approximate
\begin{equation}
  Y \simeq \slots (A + B)\,.
  \label{eq:approx_Y}
\end{equation}
If we now lean on the expression of first and second order moments of a geometric r.v. and plug  \eqref{eq:approx_Y} into \eqref{eq:age_AT-IRSA} we get, after simple manipulations
\begin{equation}
\Delta_{\textsf{at-irsa}} \simeq \frac{\slots}{2} + \frac{\slots\maxLoad + \psucc\nodes}{2\tru^*} + \frac{\slots^2}{2} \, \frac{\maxLoad -\tru^*}{\slots (1-\psucc) \tru^* + \psucc^2 \nodes}
\label{eq:age_AR-IRSA_full}
\end{equation}
where $\tru^* = \maxLoad \psucc$ is the average throughput at the target load.

\begin{figure}
  \centering
  \includegraphics[width=0.85\columnwidth]{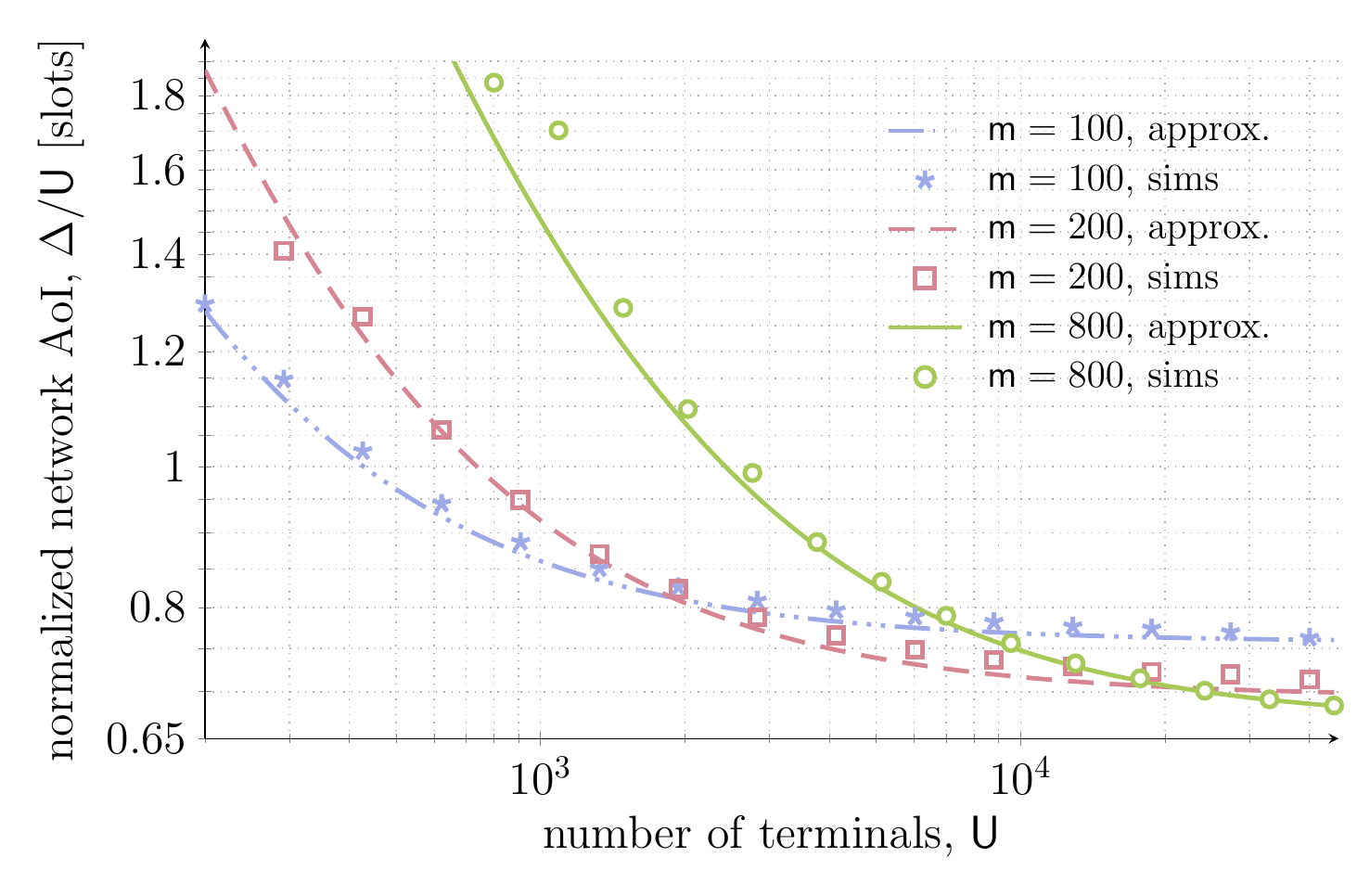}
  \caption{Normalized network AoI for \ac{AT-IRSA} vs number of terminals in the system, \nodes. Lines report results obtained via analytical apprxomation, whereas markers simulation outcomes. In all cases, $\Lambda(x)=x^3$.}
  \label{fig:analysis}
\end{figure}

The normalized network AoI for AT-IRSA estimated using \eqref{eq:age_AR-IRSA_full} is reported against the terminal population \nodes\ in Fig.~\ref{fig:analysis}, considering different frame sizes. Remarkably, an excellent match with simulation results is achieved despite the considered assumptions,\footnote{We recall that simulations implement the complete version of the scheme under a collision channel model, and do not resort to the simplifying assumptions made in the analysis for the sake of tractability.} confirming that the proposed approximation is capable of capturing the fundamental trends of the scheme in terms of AoI. From this standpoint, the simple derived formulation offers a useful tool for proper system design and protocol parameter tuning.

Furthermore, by exploring larger values of \nodes\ (up to $50$ thousands nodes), the plot suggests the existence of an asymptotic scaling law for the normalized network AoI of AT-IRSA. In other words, for $\nodes \rightarrow \infty$, a trend in the form $\Age_{\textsf{at-irsa}} = \alpha \cdot \nodes$ seems to emerge, where $\alpha$ is expected to depend not only on the considered frame size \slots, but also on the employed protocol distribution $\Lambda(x)$ \--- not explored in this paper due to space constraints. We note that such a behavior is akin to that identified in the literature for age oriented threshold-based policies applied to slotted ALOHA \cite{Bidokhti22:TIT,Uysal21_AlohaThresh,Uysal22:MiSTA}. Along this line, we conclude our discussion by comparing the age performance achieved by AT-IRSA and these state of the art benchmarks, as highlighted in Tab.~\ref{tab:comparison}. In addition to slotted ALOHA, already discussed in this section, we focus on three schemes: i) threshold ALOHA (TA), introduced in \cite{Uysal21_AlohaThresh} and resorting to fixed age-threshold and access probability values; ii) stationary age-based thinning (SAT) \cite{Bidokhti22:TIT}, which dynamically adapts the threshold based on the current congestion level; and iii) mini-slotted threshold ALOHA (MiSTA) \cite{Uysal22:MiSTA}, which refines TA by introducing an initial contention phase among nodes with high AoI. As shown in the table, AT-IRSA is capable to significantly reduce the network AoI compared to all the schemes, reaching a value as low as $0.6848 \cdot \nodes$ for the case $\Lambda(x)=x^3$ and $\slots=800$, compared, e.g., to the scaling law $0.9641 \cdot \nodes$ achieved by the best performing MiSTA scheme.

{\renewcommand{\arraystretch}{1.6}
\begin{table}
\centering
\caption{Normalized network AoI $\Age/\nodes$ for large (asymptotic) network population for different schemes: slotted aloha (SA), threshold aloha (TA) \cite{Uysal21_AlohaThresh}, stationary age-based thinning (SAT) \cite{Bidokhti22:TIT}, mini-slotted threshold aloha (MiSTA) \cite{Uysal22:MiSTA}. For AT-IRSA, $\nodes=45000$ nodes, $\Lambda(x) = x^{3}$ and $\slots=800$.}
\label{tab:comparison}
\begin{tabular}{c|c|c|c|c}
SA & TA \cite{Uysal21_AlohaThresh}	& SAT \cite{Bidokhti22:TIT}	& MiSTA \cite{Uysal22:MiSTA} & AT-IRSA \\
\hline
\hline
$e\simeq 2.7183$ & $1.4169$ & $e/2\simeq 1.3591$ & $0.9641$ & $0.6849$
\end{tabular}
\end{table}
}

\section{Concluding Remarks and Outlook}
\label{sec:conclusions}

In this paper, we have presented a variation of the IRSA access protocol \cite{Liva11:TCOM}, aiming to improve AoI performance for massive IoT networks. The scheme, named age-threshold IRSA relies on feedback provided by the receiver to dynamically select and distribute a minimum level of AoI that transmitters have to be experiencing in order to contend for the channel. By means of detailed network simulations, we have shown that the approach is capable to offer important improvements over basic IRSA. In addition, we have derived a simple analytical approach that provides a tight approximation to the average network AoI of the scheme. For large populations, AT-IRSA significantly outperforms existing threshold-based variations of slotted ALOHA studied in the literature. The promising gains we identified are meant to stimulate further research on the topic, studying, for instance, the impact of the replica distribution $\Lambda(x)$, as well as to confirm the existence of an asymptotic scaling law for the AoI of AT-IRSA.

\bibliographystyle{IEEEtran}
\bibliography{IEEEabrv,biblio_AoI,biblio_RandomAccess}

\begin{thebibliography}{10}
\providecommand{\url}[1]{#1}
\csname url@samestyle\endcsname
\providecommand{\newblock}{\relax}
\providecommand{\bibinfo}[2]{#2}
\providecommand{\BIBentrySTDinterwordspacing}{\spaceskip=0pt\relax}
\providecommand{\BIBentryALTinterwordstretchfactor}{4}
\providecommand{\BIBentryALTinterwordspacing}{\spaceskip=\fontdimen2\font plus
\BIBentryALTinterwordstretchfactor\fontdimen3\font minus
  \fontdimen4\font\relax}
\providecommand{\BIBforeignlanguage}[2]{{%
\expandafter\ifx\csname l@#1\endcsname\relax
\typeout{** WARNING: IEEEtran.bst: No hyphenation pattern has been}%
\typeout{** loaded for the language `#1'. Using the pattern for}%
\typeout{** the default language instead.}%
\else
\language=\csname l@#1\endcsname
\fi
#2}}
\providecommand{\BIBdecl}{\relax}
\BIBdecl

\bibitem{Uysal22_Semantic}
E.~Uysal, O.~Kaya, A.~Ephremides, J.~Gross, M.~Codreanu, P.~Popovski,
  M.~Assaad, G.~Liva, A.~Munari, B.~Soret, T.~Soleymani, and K.~Johansson,
  ``Semantic communications in networked systems: A data significance
  perspective,'' \emph{{IEEE} Netw.}, vol.~36, no.~4, 2022.

\bibitem{Yates19_TIT}
R.~D. {Yates} and S.~K. {Kaul}, ``The age of information: Real-time status
  updating by multiple sources,'' \emph{{IEEE} Trans. Inf. Theory}, vol.~65,
  no.~3, March 2019.

\bibitem{Ephremides20:TNET}
A.~Maatouk, S.~Kriouile, M.~Assaad, and A.~Ephremides, ``The age of incorrect
  information: A new performance metric for status updates,'' \emph{{IEEE/ACM}
  Trans. Netw.}, vol.~28, no.~5, 2020.

\bibitem{Soleymani20_valueInfo}
T.~Soleymani, J.~Baras, and S.~Hirche, ``Value of information in feedback
  control,'' \emph{{IEEE} Trans. Autom. Control}, 2020.

\bibitem{Kellerer19}
O.~Ayan, M.~Vilgelm, M.~Kl\"{u}gel, S.~Hirche, and W.~Kellerer,
  ``Age-of-information vs. value-of-information scheduling for cellular
  networked control systems,'' in \emph{Proc. ACM/IEEE ICCPS}, 2019.

\bibitem{Chiariotti22:TCOM}
F.~Chiariotti, J.~Holm, A.~Kalor, B.~Soret, S.~K. Jensen, T.~B. Pedersen, and
  P.~Popovski, ``Query age of information: Freshness in pull-based
  communication,'' \emph{{IEEE} Trans. Commun.}, vol.~70, no.~3, 2022.

\bibitem{Uysal22:ISIT}
M.~E. Ildiz, O.~T. Yavascan, E.~Uysal, and O.~T. Kartal, ``Query age of
  information: Optimizing aoi at the right time,'' in \emph{Proc. IEEE ISIT},
  2022.

\bibitem{Kaul11_Secon}
S.~{Kaul}, M.~{Gruteser}, V.~{Rai}, and J.~{Kenney}, ``Minimizing age of
  information in vehicular networks,'' in \emph{Proc. IEEE SECON}, June 2011.

\bibitem{Kaul11_Globecom}
S.~{Kaul}, R.~{Yates}, and M.~{Gruteser}, ``On piggybacking in vehicular
  networks,'' in \emph{Proc. IEEE GLOBECOM}, Dec 2011.

\bibitem{Uysal20_TIT}
Y.~Sun, Y.~Polyanskiy, and E.~Uysal, ``Sampling of the {Wiener} process for
  remote estimation over a channel with random delay,'' \emph{{IEEE} Trans.
  Inf. Theory}, vol.~66, no.~2, Feb. 2020.

\bibitem{Durisi19_JSAC}
R.~{Devassy}, G.~{Durisi}, G.~C. {Ferrante}, O.~{Simeone}, and E.~{Uysal},
  ``Reliable transmission of short packets through queues and noisy channels
  under latency and peak-age violation guarantees,'' \emph{{IEEE} J. Sel. Areas
  Commun.}, vol.~37, no.~4, 2019.

\bibitem{Shreedhar22:INFOCOM}
T.~Shreedhar, S.~Kaul, and R.~Yates, ``Coexistence of age sensitive traffic and
  high throughput flows: Does prioritization help?'' in \emph{Proc. IEEE
  INFOCOM 2022}, 2022.

\bibitem{Uysal21:ACP}
U.~Guloglu, S.~Baghaee, and E.~Uysal, ``Evaluation of age control protocol
  {(ACP) and ACP+ on ESP32},'' in \emph{Proc. IEEE ISWCS}, 2021.

\bibitem{Modiano19_TNET}
I.~{Kadota}, A.~{Sinha}, and E.~{Modiano}, ``Scheduling algorithms for
  optimizing age of information in wireless networks with throughput
  constraints,'' \emph{IEEE/ACM Trans. Netw.}, vol.~27, no.~4, 2019.

\bibitem{Ephremides19_Infocom}
A.~{Maatouk}, M.~{Assaad}, and A.~{Ephremides}, ``Minimizing the age of
  information: {NOMA} or {OMA}?'' in \emph{Proc. IEEE INFOCOM Workshops}, April
  2019.

\bibitem{Abramson77:PacketBroadcasting}
N.~Abramson, ``The throughput of packet broadcasting channels,'' \emph{{IEEE}
  Trans. Commun.}, vol. COM-25, no.~1, pp. 117--128, 1977.

\bibitem{LoRa}
{LoRa Alliance}, ``{The LoRa Alliance Wide Area Networks for Internet of
  Things},'' \url{www.lora-alliance.org}.

\bibitem{Yates17:AoI_SA}
R.~Yates and S.~K. Kaul, ``Status updates over unreliable multiaccess
  channels,'' in \emph{Proc. IEEE ISIT}, 2017.

\bibitem{Modiano18_AoI}
R.~Talak, S.~Karaman, and E.~Modiano, ``Distributed scheduling algorithms for
  optimizing information freshness in wireless networks,'' in \emph{Proc. IEEE
  SPAWC}, June 2018.

\bibitem{Yates20_ISIT}
R.~Yates and S.~Kaul, ``Age of information in uncoordinated unslotted
  updating,'' in \emph{{Proc. IEEE ISIT}}, 2020.

\bibitem{Munari22:Globecom}
A.~Munari, ``On the value of retransmissions for age of information in random
  access networks without feedback,'' in \emph{Proc. IEEE Globecom}, 2021.

\bibitem{Munari21:Balkancom}
A.~Munari and E.~Uysal, ``Information freshness in random access channels for
  {IoT} systems,'' in \emph{Proc. IEEE BalkanCom}, 2021.

\bibitem{Paolini14_CommMag}
E.~Paolini, C.~Stefanovic, G.~Liva, and P.~Popovski, ``Coded random access:
  applying codes on graphs to design random access protocols,'' \emph{{IEEE}
  Commun. Mag.}, vol.~53, no.~6, pp. 144--150, 2015.

\bibitem{Paolini15:TIT}
E.~Paolini, G.~Liva, and M.~Chiani, ``Coded slotted {ALOHA}: A graph-based
  method for uncoordinated multiple access,'' \emph{{IEEE} Trans. Inf. Theory},
  vol.~61, no.~12, 2015.

\bibitem{Stefanovic12:COML}
C.~Stefanovic, P.~Popovski, and D.~Vokobratovic, ``Frameless {ALOHA} protocol
  for wireless networks,'' \emph{{IEEE} Commun. Lett.}, no.~12, 2012.

\bibitem{Sandgren17_TCOM}
E.~{Sandgren}, A.~{Graell i Amat}, and F.~{Brännström}, ``On frame
  asynchronous coded slotted aloha: Asymptotic, finite length, and delay
  analysis,'' \emph{IEEE Trans. Commun.}, vol.~65, no.~2, 2017.

\bibitem{Clazzer18:ECRA}
F.~Clazzer, C.~Kissling, and M.~Marchese, ``Enhancing contention resolution
  {ALOHA} using combining techniques,'' \emph{{IEEE} Trans. Commun.}, vol.~66,
  no.~6, 2018.

\bibitem{Munari21_TCOM_AoI}
A.~Munari, ``Modern random access: an age of information perspective on
  irregular repetition slotted {ALOHA},'' \emph{{IEEE} Trans. Commun.},
  vol.~69, no.~6, Jun. 2021.

\bibitem{Munari21:Asilomar}
A.~Munari, F.~Lazaro, G.~Durisi, and G.~Liva, ``An age of information
  characterization of frameless {ALOHA},'' in \emph{Proc. IEEE Asilomar}, 2021.

\bibitem{Ngo21:Asilomar}
K.-H. Ngo, G.~Durisi, and A.~G. i~Amat, ``Age of information in prioritized
  random access,'' in \emph{Proc. IEEE Asilomar}, 2021.

\bibitem{Uysal21_AlohaThresh}
O.~T. Yavaskan and E.~Uysal, ``Analysis of slotted {ALOHA} with an age
  threshold,'' \emph{{IEEE} J. Sel. Areas Commun.}, vol.~39, no.~5, May 2021.

\bibitem{Uysal22:MiSTA}
M.~Ahmetoglu, O.~T. Yavascan, and E.~Uysal, ``{MiSTA}: An age-optimized slotted
  {ALOHA} protocol,'' \emph{{IEEE} Internet Things J.}, vol.~9, no.~17, 2022.

\bibitem{Bidokhti22:TIT}
X.~Chen, K.~Gatsis, H.~Hassani, and S.~Bidokhti, ``Age of information in random
  access channels,'' \emph{{IEEE} Trans. Inf. Theory}, vol.~68, no.~10, 2022.

\bibitem{Liva11:TCOM}
G.~Liva, ``Graph-based analysis and optimization of contention resolution
  diversity slotted {ALOHA},'' \emph{{IEEE} Trans. Commun.}, no.~2, 2011.

\bibitem{dvbrcs2}
{ETSI}, ``{EN 301 545-2: Digital Video Broadcasting (DVB); Second Generation
  DVB Interactive Satellite System (DVB-RCS2); Part 2: Lower Layers for
  Satellite standard},'' Tech. Rep., 2014.

\end{thebibliography}

\end{document}